\documentclass[a4paper,11pt]{article}
\usepackage[cp866]{inputenc}
\usepackage[dvips]{graphicx}
\usepackage{epsfig}
\usepackage[russian]{babel}
\begin{document}
\Large

\begin{center}

\section*{Optical Multicolor Observations of the  SS 433=V 1343 Aql Microquasar}

 2010  A.N. Sazonov

Sternberg Astronomical Institute, Universitetskii pr. 13, Moscow,
119992 Russia

\end{center}

\quad

\textbf{Abstract}

~ We report BVR photometry of the V1343 Aql= SS 433 microquasar at
different phases of the 13--day orbital cycle for the 1986--1990
observing seasons. The data include five complete cycles of the
163$^{d}$ precession period of the system. We obtain mean light
curves and color--color diagrams with the orbital period for all
intervals of precession phases. The optical component of the close
binary system (CBS) fills its critical Roche lobe and loses mass
on the thermal relaxation time scale. Gaseous flow show up
actively in the system and activity manifestations differ
substantially at different precession phases.

~ The collimated relativistic jets perpendicular to the plane of
the disk appear to be associated with supercritical accretion onto
the compact relativistic object in the massive CBS, which shows up
in the shapes of the light curves at different orbital and
precession phases. An analysis of color indices confirmed the
earlier discovered peculiarities of the system
~\cite{Shakhovskoi1996}:

(1) A  "disk corona" around the compact object.

(2) Phase shifts between orbital light curves and different
heights of light maxima for different passbands and at all phases
of the  163$^{d}$ precession period.

\quad

\textbf{Introduction}

 The unique astrophysical object SS~433 (V1343 Aql), $\alpha =
19^h11^m49^s.56$, $\delta +04^\circ58'57''.6$ (J2000), has been
for many years attracting attention of observing astronomers as
well as theoretical physicists all over the world. The
evolutionary status of the close binary system SS~433 is the X-ray
stage of massive close binary systems (MCBS). The active state of
these stars, which are at the X-ray stage of their evolution, is
explained by a spectrum of changes in the process of accretion of
material on the CBS compact object. However, SS~433 is strongly
distinguished among other X-ray sources by that in the X-ray range
radiate X-ray jets, not the accretion disk~\cite{Ctewart1989}.

 The high rate of accretion on the compact object ($\sim 10^{-4}$
M$_{\odot}$/year) produces these jets aligned with the accretion
disk axis. The velocity of gas in the jets is $\sim
80,000\textrm{~км/с} = 0.26c$. The orbital period of the system
is $13.082^d$, and the precession period of the disk and jets is
$162.5^d$ ~\cite{Margon1984}, ~\cite{Cherepashchuk1989}.

 The main minimum of the orbital light curve corresponds to the
eclipse of the accretion disk by the normal star. In this CBS the
optical star is at a later evolutionary stage. It overfills its
critical Roche lobe and outflows onto the relativistic object on
the thermal relaxation timescale at a rate of $\sim 10^{-4}$
M$_{\odot}$/year. This results in supercritical accretion onto
the relativistic object ~\cite{Shakura1973}, because the
appearance of collimated relativistic outflows of material from
the central parts of the thick accretion disk is a novel and
unexpected feature of the supercritical accretion mode.

 Simultaneous measurements of brightness in several spectral bands
will yield unbiased information for undistorted colors of the
object SS~433 and for updating existing models of it.

\quad

\textbf{Observations}

 Systematic photoelectric \emph{BVR} observations (\emph{W} band
observations are also available) were carried out by the author in
1986--1990. They are listed in Table~1, which can be found
at\mbox{({http://lnfm1.sai.msu.ru/$\sim$ sazonov/V1343 Aql=SS
433})}.

 In these observations the mean errors in the \emph{BVR} spectral
bands were from 0$^{m}$.010 to 0$^{m}$.035.

 In total we have obtained 1129 individual measurements in 295
nights during five years of observations. The observations were
carried out with single--channel \emph{WBVR} photometers on the
following instruments: 600--mm reflector of the Crimean Laboratory
of the Sternberg Astronomical Institute (Ukraine, altitude above
sea level 550~m) with a $27^{\prime\prime}$ diaphragm in the
\emph{WBVR} instrumental photometric system; Zeiss--600 reflector
on Mt. Maidanak (Uzbekistan, altitude above sea level 2600~m);
480--mm reflector of the Trans-Ili Alatau Observatory (Kazakhstan,
altitude above sea level 2760~m).

 We have used as a radiation detector a FEU-79 photomultiplier tube
(S-20 multi-alkali photocathode). The broadband \emph{WBVR}
system has been used owing to its better determinacy. The system
was described by \cite{Straizys1977}.

\quad

\textbf{Reduction to the standard system}

 In electrophotometric observations of great importance is proper
matching of instrumental corrections to the standard Johnson
photometric system. In joint observations with T.R.~Irsmambetova
at the Mt.Maidanak Observatory in 1987--1989 the author used the
reduction coefficients of the instrumental system in the work on
the Zeiss-600 telescope (single-channel \emph{WBVR}
electrophotometer with automatic control system; as a radiation
detector a FEU-79 photomultiplier tube was used).

 The relative spectral sensitivity of the system was rather
stable. It was checked twice in the observational season
(spring--summer and summer--autumn).

 The coefficients of reduction to the standard photometric system
were obtained from repeated measurements of standard stars in the
areas SA 107, 108, 111--113 \cite{Moffett1979}.

 We have the following quantities:

 $$
 B-V=1.071(b-v)-0^m.068
$$
$$
(\pm 0.021)   (\pm0.018)
$$

$$
V-R=0.803(v-r)+0^m.173
$$
$$
(\pm0.033)   (\pm 0.014)
$$

instrumental \emph{bvr} stellar magnitudes; \emph{BVR} stellar
magnitudes in the Johnson photometric system.

 To calculate corrections to the determined stellar magnitudes
from the known instrumental color indices we use the following
expressions:

$$
B-b=0.094(b-v)-0^m.099
$$
$$
V-v=0.014(b-v)-0^m.015
$$
$$
R-r=0.234(v-r)-0^m.208.
$$

 In subsequent observations of 1991--1998 (including those at
other observatories) we derived the coefficients of reduction to
the standard Johnson photometric system from measurements of
standard stars in the areas of h and $\chi$~Per (NGC~869, the
author measured 12 standard stars) and NGC~884 (13 standard
stars).

 It should be noted that, owing to the extremely faint brightness
of the object SS~433, short integration time of the useful signal,
$60^s-90^s$, and small aperture of the telescopes used
(Zeiss--600), the \emph{W}-band data were obtained only at the
instants of the maximum light of the system (T3) with a temporal
resolution of 5--6~min on a time interval of 3--4~h (more seldom
$\sim$5~h).

 On medium-aperture telescopes (1200~mm; e.g., AZT-11 of the
Crimean Astrophysical Observatory) we had rms errors of the
measurements estimated from pulse statistics with regard to
background at the periods of maximum light 0$^m.045$, 0$^m.035$,
0$^m.020$, 0$^m.015$, 0$^m.015$ in the \emph{UBVRI} bands,
respectively.

 On small instruments (with an aperture of 480--600~mm) in the
\emph{BVR} bands we had rms errors estimated from pulse
statistics with regard to the background in the periods of
maximum brightness 0$^m.040$, 0$^m.020$, 0$^m.008$ in the
\emph{B, V, R} bands. The mean errors of the estimated brightness
can reach 0$^m.08$ in the \emph{W} band, 0$^m.04$ in the \emph{B}
band, and do not exceed 0$^m.03$ in the \emph{V, R} bands.

\quad

\textbf{Ephemeris}

 In this work we have used photometric elements of the middle of
eclipses (phases $\phi$) and maxima of the out-of-eclipse
brightness (phases $\psi$) taken from ~\cite{Wittone1983},
~\cite{Shakhovskoi1996}, respectively:

\textbf{Min I
Hel=2444332}$^{d}$\textbf{.98+13}$^{d}$\textbf{.086}$^{.}$\textbf{E
(phasa $\phi$)}

(phase $\phi = 0$ corresponds to the instant of the occultation of
the accretion disk by the ``normal'' star) and

\textbf{Max=T3=2443666}$^{d}$\textbf{.29+163}$^{d}$\textbf{.34}$^{.}$\textbf{E
(phasa $\psi)$}

(where T3 is the instant of the maximum separation of the moving
emission lines).

It is accepted that at this instant $\psi = 0$.

 All photometry was done with respect to the comparison star C1 =
Kemp1~\cite{Kemp1986} and reference star B as
in~\cite{Shakhovskoi1996}:

(C1=GSC 0471-01564 $19^h11^m47^s.66+5^\circ00'35''.0 (J2000)$);

(B=GSC 0471-00142 $19^h11^m47^s.04+4^\circ59'38''.1 (J2000)$),

 The star C1 was linked to standards from the list
of~\cite{Neckel1980}:

$U$=$14^m.22$; $W$=$14^m.14$; $B$=$12^m.93$; $V$=$11^m.46$;
R=$10^m.09$.

 For the reference star B we took magnitudes:

 $B$=$14^m.57$; $V$=$13^m.46$; $R$=$12^m.52$.

\quad

\textbf{Interpretation of the observations}

 All observational data reported in this paper as well as their
interpretation were compared with similar works of other authors
for the last 30~years:

~\cite{Shakhovskoi1996}, ~\cite{Cherepashchuk1989},
~\cite{Cherepashchuk1991}, ~\cite{Irsmambetova1997},
~\cite{Aslanov1987},~\cite{Glad1987}, ~\cite{Cherepashchuk1982},
~\cite{Irsmambetova2001}, ~\cite{Goranskii1998}, ~\cite{Glad1983},
~\cite{Catchpoll1984}, ~\cite{Efimov1984}, ~\cite{Dolan1997}.

 Optical light curves of the close binary system (CBS) V1343 Aql =
SS 433 obtained in B, V and R bands for the period 1986-1990 are
qualitatively similar (Fig. 1a, 1b and 1c).

 The average depths of the primary minimum (when the accretion disk
around the tight companion is occulted by the "normal" star) for
this period is 0$^m.65-0^m.75$, 0$^m.50-0^m.60$ and
0$^m.35-0^m.45$ mag in $B$, $V$ and $R$--filters, respectively.

 The light curves evolve during the precession cycle of the system
which may be attributed to the gas streams activity in the CBS and
their interaction with the "floating" accretion disk which spatial
orientation drive the direction of the relativistic jets.

 Phase shifts of the orbital light curves and variation of the
maxima heights confirm the presence of asymmetry of the brightness
distribution in the accretion disk. They also support the fact
that the backside of the disk (with respect to the orbital motion)
is more luminous which was first noticed in [1].

 The variations of color indices with the orbital and precession
periods are well confirmed for Min I and Min II by the above
mentioned observation seasons (see, for example, the 1988 data in
Fig. 2a, 2b, 2c and Fig. 3a, 3b, 3c).

 Analyzing the photometric data, we should pay special attention
 to comparitive and qualitative analysis with early works of the
 author.

The small amplitude of chanzing of color index $B-V$ is clearly
seen on the diagrams of color indexes.

Evidently this is one of the consequences of relatively small
difference in the $B-V$ color of thermal continuum, radiated by
the bright optical star and different parts of accretion disk in
the system [1], because at temperature higher an 20000 K the color
$B-V$ of optical radiation practically does not depend on the
temperature.

We should note that the character of periodic brightness
variations of SS 433 in these observing seasons is in qualitative
agreement vith the date from othei authors for the period
1978--1990.

This allows us to conclude about some stability of regular
brightness variations with periods $163^d$ (precession period,
amplitude $\Delta V=0^m.75\div 0^m.85$), $13^d.086$ (orbital
period, amplitude $\Delta V=0^m.80\div 0^m.90$).

In the observing period 1986--1990 a the series 5--7 flares were
observed each observing season (the object was especially active
in 1988) with amplitude frow $\sim 0^m.30$ to $\sim 0^m.50$ in the
$V$ band.

The flares of optical radiation of SS 433 in these observing
seasons are distributed nearly evenly on all the phases of
precession period.

 Undoubtedly, the amplitude and shape of the orbital light curve
of the system is functionally dependent on the precessional phase
of the appearance of the relativistic jets in space. This is well
visible from the comparison of orbital light curves for different
observational seasons (see, e.g., 1986 and 1988) at the same
precessional phases.

 In these observations rms errors estimated from pulse statistics
with regard to the background in periods of maximum brightness
were 0$^{m}$.040, 0$^{m}$.020, 0$^{m}$.008 in the \emph{B, V, R}
bands, respectively. Mean errors of brightness estimates may
reach $0^{m}.08$.

 In the regular precessional variability the out-of-eclipse CBS
brightness also changes with the precessional phase, and at
instant T3, $\psi = 0.0$ (maximum separation of the relativistic
emission lines) the brightness is maximum ,whereas near $\psi =
0.5$ the brightness is minimum. This is visible rather clearly in
all graphs of this observational interval.

 These, relatively regular variations are superimposed by
unpredictable bursts of brightness, which are caused by the
active state of the object (these properties of the object
persisted during all the years of the author's observations). The
brightness increase during these bursts is of the order of
$\Delta V_{max}=1^{m}.2$.

 The largest fluctuations of brightness take place at instant T3
($\psi = 0.00$), which corresponds to the maximum opening of the
disk toward the observer. Near the phase $\psi = 0.50$, when the
disk turns its other pole to the observer, the scatter of the
observed points becomes minimum; this is well visible from the
lower envelope of the light curve. This is especially notable in
the seasons of 1987 and 1989.

 In these seasons the accuracy of the photometric maxima (T3) was
not very high because of substantial distortions of the orbital
light curves due to the activity, which may last up to 90 days
and longer. The upper level of the light curve is more rarefied,
and it is formed by separate flares.

\quad

\textbf{The system's flare activity}

 On the basis of long-term photometric observations and of their
analysis, many authors have established that the object SS~433
spends in the active state more than $20\%$ of the time. In the
optical range its flare activity manifests itself differently.
Sometimes series of flares are observed, and sometimes isolated
flares appear with a \emph{V} band amplitude of $\sim 0^{m}.8$ and
with varying duration.

 A series of flares can last up to 40\% of the precessional
period. It results in an appreciable chaos in the light curve
geometry. Similar phenomena were observed, e.g., in 1979, 1980,
1982, and 1986 (literary data) as well in the observations of
1988 (this work).

 In these observational seasons the optical flares of SS~433 are
distributed almost uniformly over all phases of the precessional
period, as it is visible from the entire observational database
reported here.

 Note prolonged flares recorded in these observational seasons: ~

\textbf{The system's flare activity in 1986 was detected during:}

JD 2446619, JD 2446625, JD 2446652, JD 2446715.

\textbf{The system's flare activity in 1987 was detected during:}

JD 2446995, JD 2447016, JD 2447062, JD 2447064, JD 2447082,

JD 2447094, JD 2447096.

\textbf{The system's flare activity in 1988 was detected during:}

JD 2447318, JD 2447423, JD 2447327, JD 2447345, JD 2447360.

 The flare at observing date JD 2447434 has the following parameters:
0$^{m}.758$ in $B$-band, 0$^{m}.512$ in $V$-band, 0$^{m}.252$ in
$R$-band. The $\varphi=0.992$, $\psi=0.068$ in this case.

\textbf{The system's flare activity in 1989 was detected during:}

JD 2447712, JD 2447721, JD 2447734, JD 2447747, JD 2447748,

JD 2447798, JD 2447800.

\textbf{The system's flare activity in 1990 was detected during:}

JD 2448105, JD 2448190--JD 2448191.

\quad

\textbf{The star's nutation light curve}

 We also analyze the nutational light curve of the object based on
optical observations made in 1980--1990 (see paper
~\cite{Shakhovskoi1996}):  we first subtracted the precessional
and orbital variations and phased the residual with the  6.28-day
nutational period (Fig. 4a, 4b, 4c). The nutational light curve
qualitatively resembles a similar light curve from
~\cite{Goranskii1998}. We adopted the ephemeris for the maxima of
the nutational variations from ~\cite{Goranskii1998a}.

 The nutation phenomenon manifests itself as wobble of the
relativistic jets in the system.

 Nutation is the third reliably established period in the optical
light variations of SS~433. It is accepted that, irrespective of
the CBS precessional orientation in space, total eclipses of the
accretion disk by the optical star are never observed in SS~433.

 All these regularities testify that in the CBS SS~433 there are
partial eclipses of the precessing accretion disk by the
``normal'' star.

 From the analysis of the five--year photometric database we obtain
that the nutational period remains stable within the errors of the
observations and mathematical processing.

 In the interpretation of the observational data of the seasons of
1986--1990 (together with unpublished author's data of 1991--1994
and 1996--1998) attention is drawn to some data obtained near
precessional phases $\psi =0.00\pm 0.16$ (T3, the instant of the
maximum separation of the relativistic emission lines), which
have found no satisfactory explanation in the framework of
generally accepted other authors' scenarios of the behaviour of
the CBS SS~433 = V1343~Aql. In this paper the author attempted to
explain the above-mentioned observational features of the system.

 At precessional phases $\psi = 0.02-0.16$ (and different orbital
phases $\phi$) there are observed points at which the brightness
has an increased amplitude (but they are not flares, because the
points relax rather quickly to the mean brightness) as compared to
neighboring points within a relatively short time interval and in
brightness.

 As a rule, we chose good photometric nights. We selected
statistical photometric data in three (more seldom in four)
spectral bands to refine the already existing correlations.

\quad

\textbf{Special points on the light curves}

 Some researchers attribute these features to the nonstationary
nature of the activity of  SS 433 and to the strong variability of
the mass-loss rate in relativistic jets. However,  inflection
points are by no means the only singularities to be found on
radial-velocity curves. Other possible singular points include,
e.g., points of self-intersection of gas jets in the system
~\cite{Sazonov1988a}.

 Sazonov and Shakura ~\cite{Sazonov1988a} already pointed out the above feature
of particle trajectories in the gas flow in a close binary system
- i.e., their tendency to intersect (even in the
celestial-mechanics approximation). The intersection of ballistic
trajectories of flow particles may prove to be one of the factors
of the formation of irregularities in the gaseous jet. The
formation of irregularities may may, in turn, result in the
scintillation of the "hotspot" that develops where the jet meets
the disklike envelope of the relativistic object of the close
binary system or where it intersects with the neighboring
trajectories of the jet flow. Possible cases include the the
collimated jet trajectory portions located near time T3 (or,
rather, at phases $\varphi$=0.03-0.15).

 The authors of papers ~\cite{Bisikalo1994} and ~\cite{Bisikalo1996}
and review ~\cite{Lubow1993} point out various instabilities of
the gas flow in the vicinity of the compact object in the CBS and
indicate that the gas-flow instability shows up, among other
things, as quasi-periodic oscillations of the mass accretion rate
M and the rate of change J of the angular momentum of the matter
located near the secondary component of the CBS. The above authors
used the computed density, velocity, and temperature fields to
calculate the emission profiles of the H??? line, and found the
wings of these lines to form near the accretor and their
broadening to be determined by the high velocities of the gas flow
in the accreting matter and disk. The gas velocities near the
accretor were found to exceed the average gas velocity in the
system by a factor of more than 10 ~\cite{Bisikalo1996}. This
result agrees well with the interpretation of our optical
observations.

 Furthermore, we also established in our earlier papers that gas in
the jet is concentrated in individual clouds ~\cite{Murdin1980},
~\cite{Shklovskii1981},~\cite{Grandi1982}, in line with the above
scenario that explains the observed data points with a somewhat
excessive amplitude at precession phases $\psi$=0.02-0.16 (and
various orbital phases $\varphi$) of SS 433.

\quad

\textbf{Rapid variations of the object}

 We also performed observations on time scales corresponding to rapid
variability, ranging from 90 to 180 s per single exposure. We
studied the object over time intervals ranging from 1 to 2.5
hours. The authors of ~\cite{Chakrabarti2005} concluded, based on
the results of x-ray, optical, and radio observations made with a
temporal resolution of 16 s, that individual clouds show up at all
wavelengths.

 The existence of fast variability in the brightness of the object
SS~433 on timescales from several minutes to 20--30~min was
established and confirmed in ~\cite{Kopylov1986},
~\cite{Goranskii1987}, ~\cite{Zwitter1991}. At the present stage
of the study of this object we need high - precision photometric
data for updating the existing model of the fast variability and
for revealing still unknown mechanisms of its origin.

 The observed brightness difference in all spectral bands within
the same observational night can be explained physically by fast
variability with a timescale of 10--30~min and amplitude from
$0^m.1$ to $0^m.3$ ~\cite{Goranskii1998}, ~\cite{Goranskii1987},
~\cite{Sazonov1988}.

 Parallel photometric and spectral observations of the research
team of the Special Astrophysical Observatory ~\cite{Kopylov1986}
have shown that the fast photometric and spectral variability on
timescales $\sim$10--30~min is observed in the quiescent state of
the system at any phase of the orbital and precessional cycles.

 According to the observations in these seasons the main minimum
in the \emph{B} band is deeper than in the other bands,
especially at precessional phases $\psi =$ 0.3, 0.4, 0.5, 0.6,
and 0.7.

 The observations of the season in the mode of the fast variability
on timescales from 2 to 3--4 min (the exposure in the \emph{BVR}
or \emph{WBVR} bands) have confirmed that the dependences of $(B -
V)$ on $B$ and $(V - R)$ on $V$ are fulfilled unambiguously: the
$(B - V)$ and $(V - R)$ color indices decrease with increasing $B$
and $V$ brightness, respectfully.

 The authors of ~\cite{Kopylov1986} have drawn a conclusion about
the causes of fast variability as a result of the passage of the
relativistic jets through the circumbinary envelope. At short
timescales the fast variability is explained by the discrete
character of the jet, which consists of separate ionized gas
clouds.

 Rapid variability was observed in all photometric bands
(variations were found in the BVR data obtained by the author of
this paper and in the  $UBVRI$ data obtained in
~\cite{Shakhovskoi1996}). The authors of ~\cite{Shakhovskoi1996}
found that orbital light variations exhibit phase shifts and the
heights of the maxima (amplitudes) differ for different passbands
and different phases of the 163 - day period.

 The amplitude of physical intranight variability of  SS 433 may
amount to about $0^m.25$ mag or more (see plots for the 1988
season).

 The object shows persistent light  fluctuations in the  $R$ and $I$ bands
(~\cite{Shakhovskoi1996} and this paper) on  time scales of 60 -
90 s. Recent coordinated optical and x - ray observations found
such fluctuations in the optical $R$ band on time scales of about
10 s as a result of relatively ~\cite{Revnivtsev2004}.

 The outbursts that occur rather frequently in the system studied produce
considerable distortions in the orbital light curves observed
during every season. The 1988 season is especially remarkable in
this regard.

 According to published data, both relativistic hydrogen lines and stationary
lines exhibit rapid variations. Relativistic hydrogen lines have a
complex multicomponent structure, which varies from night to
night. The objects shows variations in the $R$ band (see
~\cite{Shakhovskoi1996} for similar observations in the $R$ and
$I$ bands). The variations involved brightness increases amounting
to $0.45^m$ and  $0.35^m$ in the $R$ and $I$ bands, respectively,
over a 10 - minute long observation (during the same night).

 Such variations were recorded several times in $UBVRI$ - (1980 - 1986)
and $WBVR$ - band (1986 - 1990) observations: JD 2445249, JD
2446302, JD 2446995, JD 2447082, JD 2447096, JD 2448053. According
to the classification proposed in ~\cite{Goranskii1998}, the
"blue"{} component of emission was present in this case with a
color index of V - R=1.9, whereas the "red"{} emission component
was absent.

\quad

\textbf{Conclusions}

 Our interpretation of the data set obtained during the entire
1986 - 1990 observing period leads us to the following important
conclusions:

1. One of the new results of this work is that we obtained
simultaneous and homogeneous multicolor observations of the object
in the $WBVR$ photometric bands. We obtained the orbital light
curves of SS 433 at the precession phases of $\psi$=0.0 - 0.1,
$\psi$=0.1 - 0.2, $\psi$=0.2 - 0.3, $\psi$=0.3 - 0.4, and
$\psi$=0.4 - 0.5, thereby extending substantially our cycle of
works that we began earlier
~\cite{Shakhovskoi1996},~\cite{Sazonov1988a}.

2. The manifestations of the activity of the system differ rather
significantly at different phases of the 163 - day precession
period.

3. During all the observing years the light curve exhibited
persistent Min I and Min II, which have the standard depths
typical for the precession phase considered in all four - to -
five photometric bands observed: the minima do not go below
$16^m.779$, $15^m.250$, and $12^m.332$ in the $B$, $R$, and $R$
passbands, respectively (in my other papers I report observations
made in the $U$ and $W$ bands using intermediate telescopes
~\cite{Shakhovskoi1996},~\cite{Sazonov1988a},~\cite{Sazonov2006}).

4. At these times the primary eclipses Min I and Min II appear as
rather sharp and deep dips seen against somewhat excessive
brightness. The light curve is somewhat asymmetric - by about
0.006 periods - with respect to the theoretical  Min I at time T3.
In the quiescent state the orbital light curve covers a single
wave per period at this phase.

5. The average magnitude of the system is somewhat brighter than
its usual level, by $0.3^m - 0.45^m$.

6. The observed phase shift of the $BVR$ light curves (and also in
the $U$ and $I$ - band light curves obtained in 1996) whose
magnitudes and signs differ at different precession phases: the
eclipse in long - wavelength bands occurs later than in short -
wavelength bands ($\psi$=0.10 - 0.12), i.e., first the cooler
(dark) regions of the accretion disk are eclipsed, and only then
hotter and brighter regions; however, the situation is reversed at
the $\psi$=0.40 - 0.45 phase:

6.1. We thus have well - defined phase shifts in the light curves
in all the photometric bands observed ($UBVRI$), which are
unambiguously related to wavelength.

6.2. Moreover, the maxima (Min I, Min II) are clearly asymmetric
at precession phases $\psi$=0.10 and $\psi$=0.60, and this
asymmetry is especially apparent in the $B$ and $V$ bands.
 It also follows from the asymmetry of the light maxima that
at the precession phase of $\psi$=0.10 (and, less conspicuosly,
but clearly enough, at the phase of $\psi$=0.60) the brightness of
the leading part (with respect to orbital motion) of the accretion
disk is lower than that of the trailing part.

6.3. Both these features (6.1-6.2) conclusively indicate that the
accretion  disk of the primary component of the CBS is irregular
and asymmetric, and, more generally, that so is the structure of
the accretion formation of the entire SS 433=V1343 Aql system.

7. The (W-B), (B-V), (V-R), and (R-I) color indices are
functionally dependent on $W,B,V$, and $R$-band magnitudes. These
linear relations strictly obey the pattern: color index decreases
with increasing brightness.

8. The source of rapid variations in SS 433=V1343 Aql does not
disappear in the system's activity stage and is not eclipsed.

9. The brightness amplitude increases towards shorter wavelengths
for all Kinds of variations of the binary (orbital, precession,
nutation variations, rapid variations within a night of
observations).

\newpage
\begin{figure}[b]
\centerline{\epsfig{file=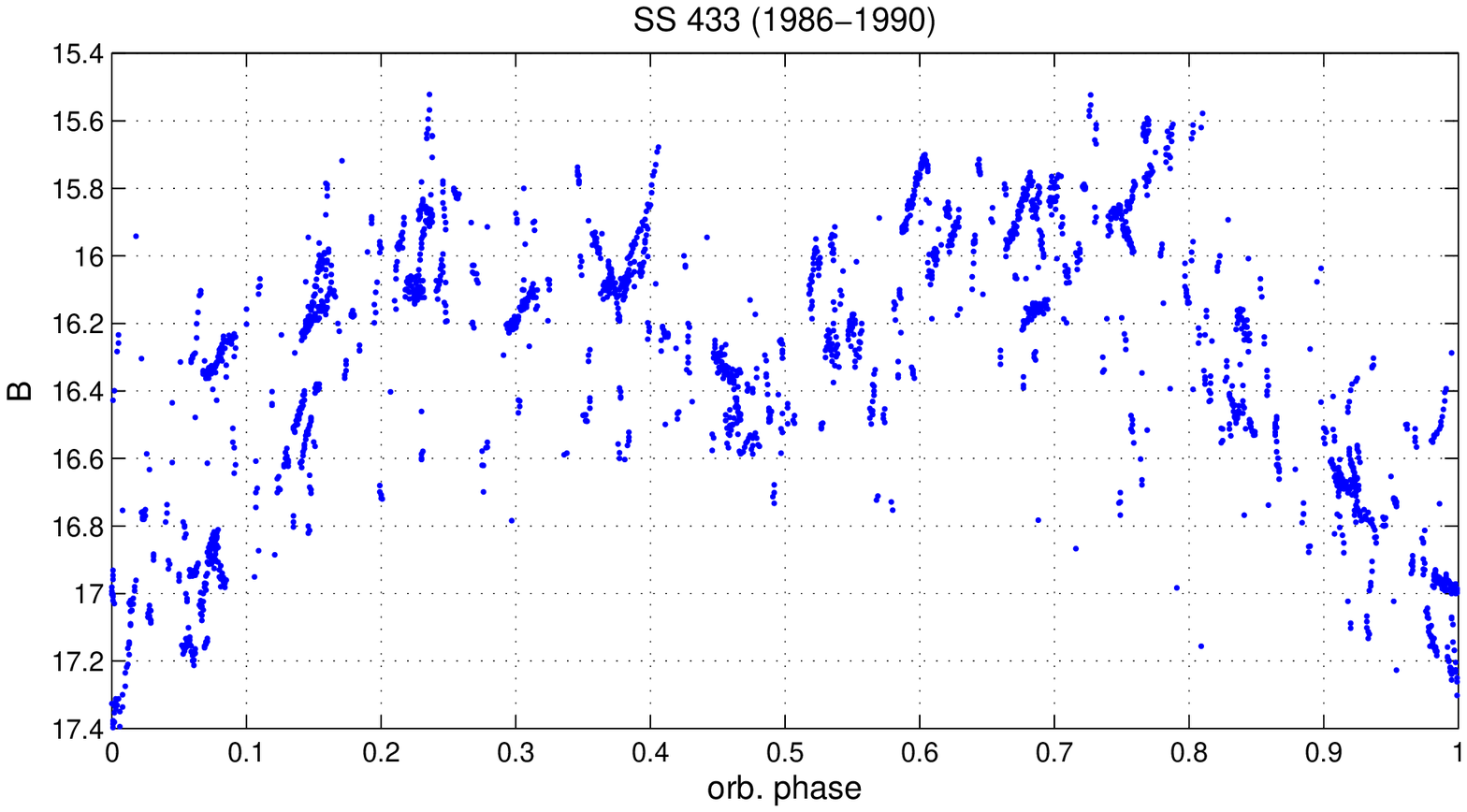,width=150mm}}
\caption{The 1986-1990 B-band light curve folded with the period
$P=13^d,086$}
\end{figure}

\newpage
\begin{figure}[b]
\centerline{\epsfig{file=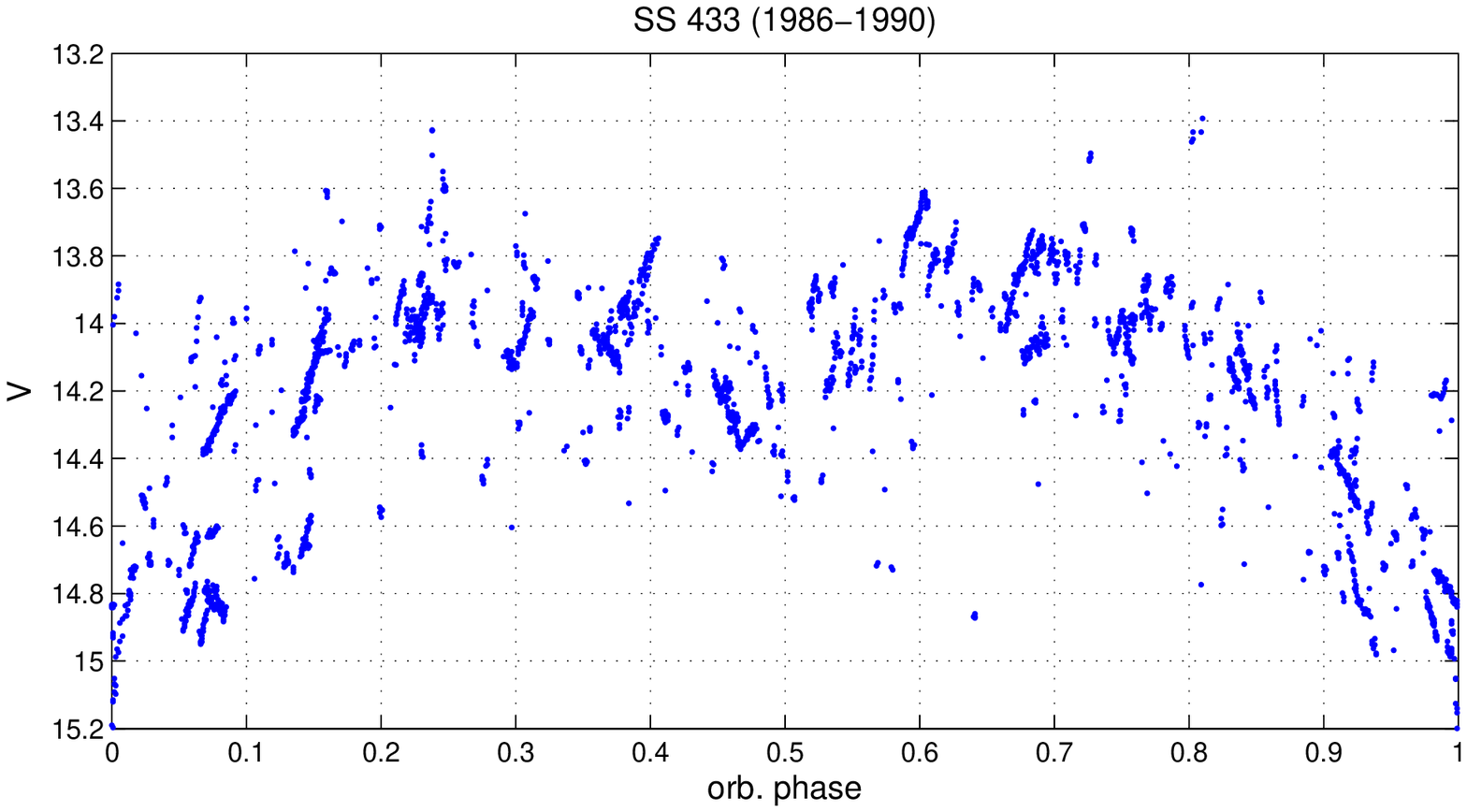,width=150mm}}
\caption{The 1986-1990 V-band light curve folded with the period
$P=13^d,086$}
\end{figure}

\newpage
\begin{figure}[b]
\centerline{\epsfig{file=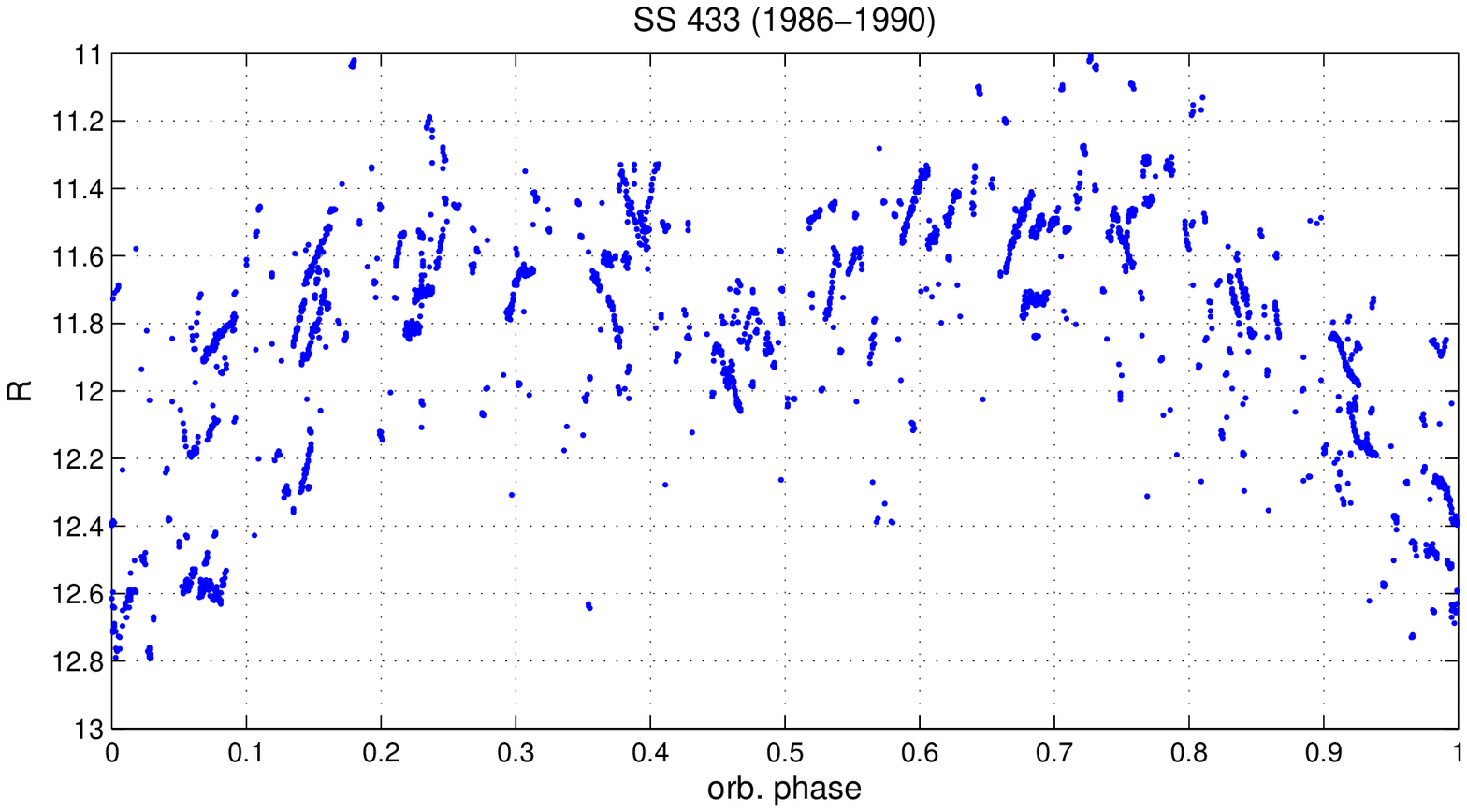,width=150mm}}
\caption{The 1986-1990 R-band light curve folded with the period
$P=13^d,086$}
\end{figure}

\newpage
\begin{figure}[b]
\centerline{\epsfig{file=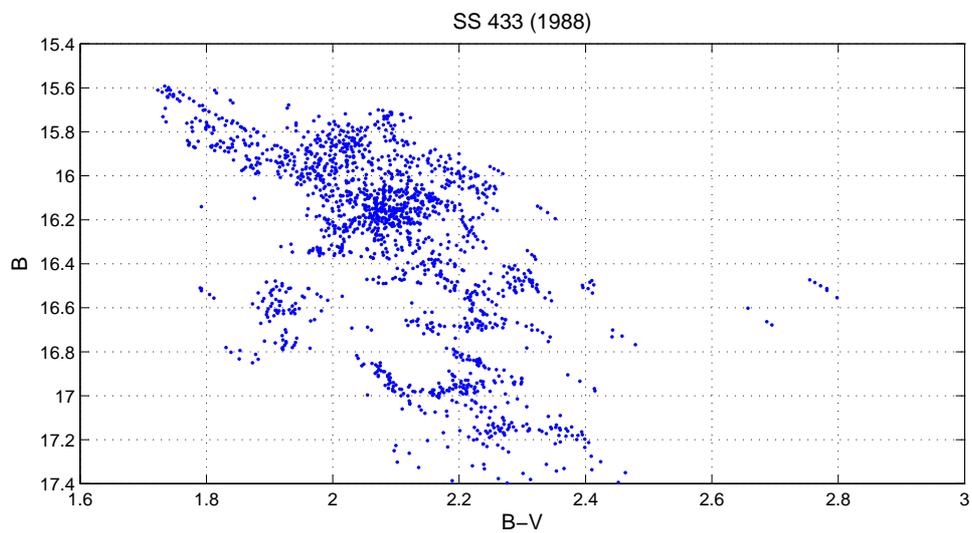,width=150mm}}
\caption{The 1986-1990 B--V color of SS 433 as a function of
B-mag}
\end{figure}

\newpage
\begin{figure}[b]
\centerline{\epsfig{file=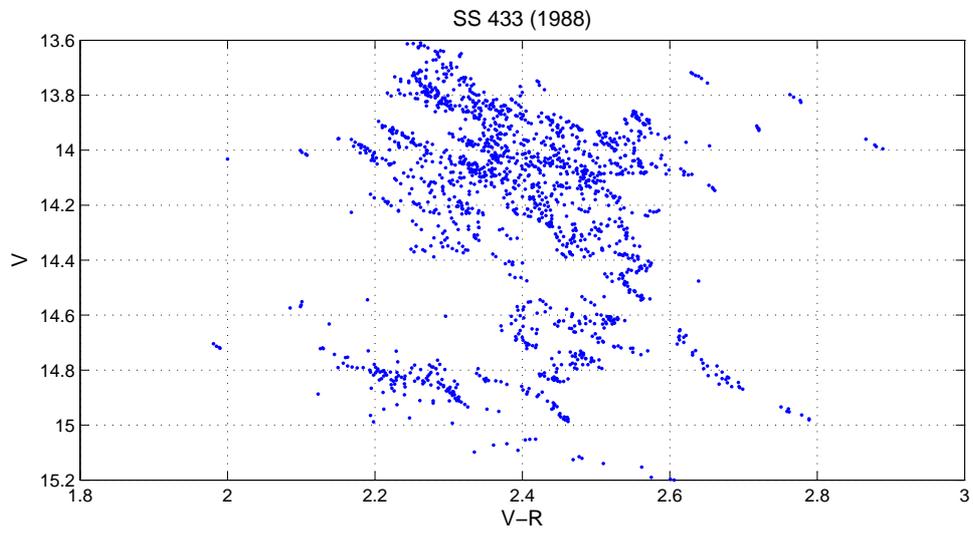,width=150mm}}
\caption{The 1986-1990 V--R color of SS 433 as a function of
V-mag}
\end{figure}

\newpage
\begin{figure}[b]
\centerline{\epsfig{file=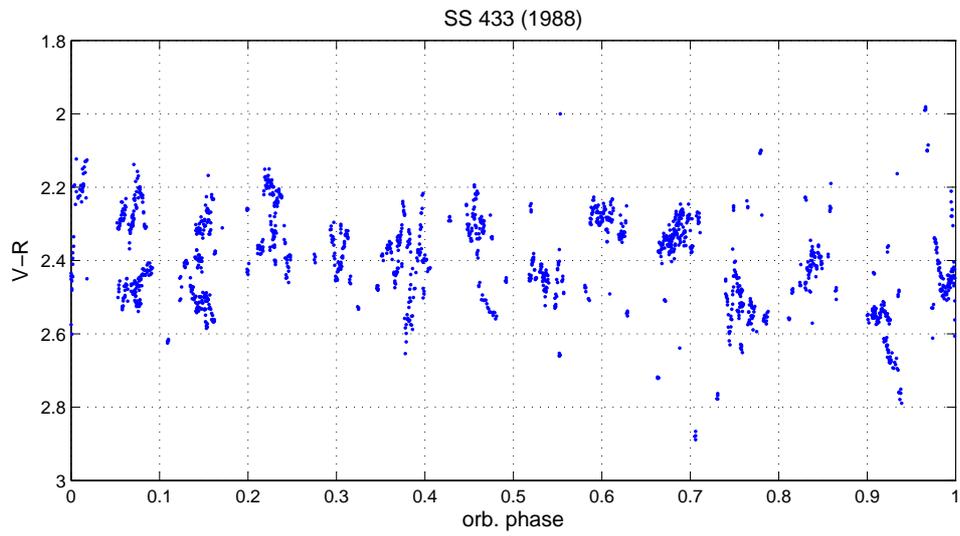,width=150mm}}
\caption{SS 433 (1988): V-R as a function of the orbital phase}
\end{figure}

\newpage
\begin{figure}[b]
\centerline{\epsfig{file=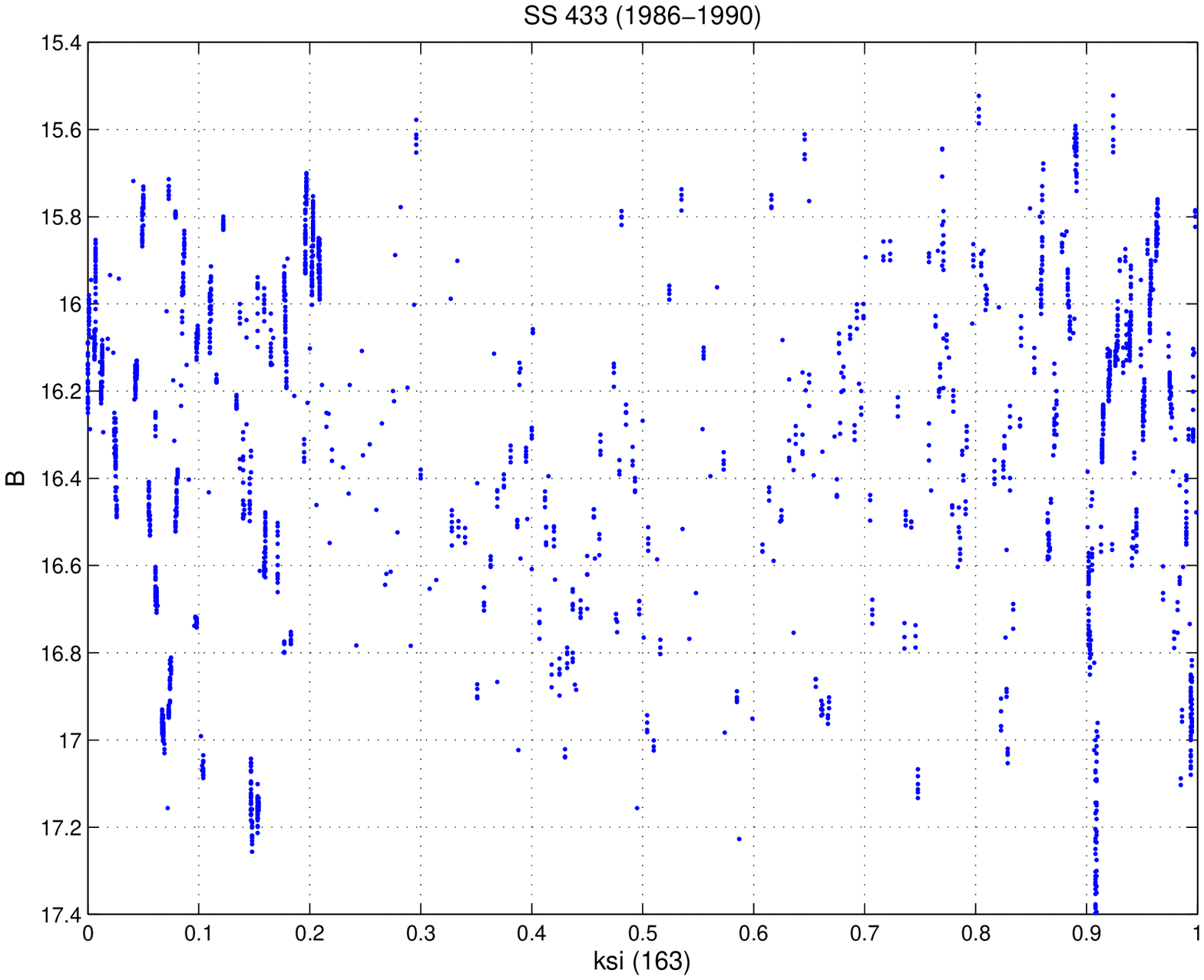,width=150mm}}
\caption{The 1986-1990 B-band light curve folded with the
precession phase}
\end{figure}

\newpage
\begin{figure}[b]
\centerline{\epsfig{file=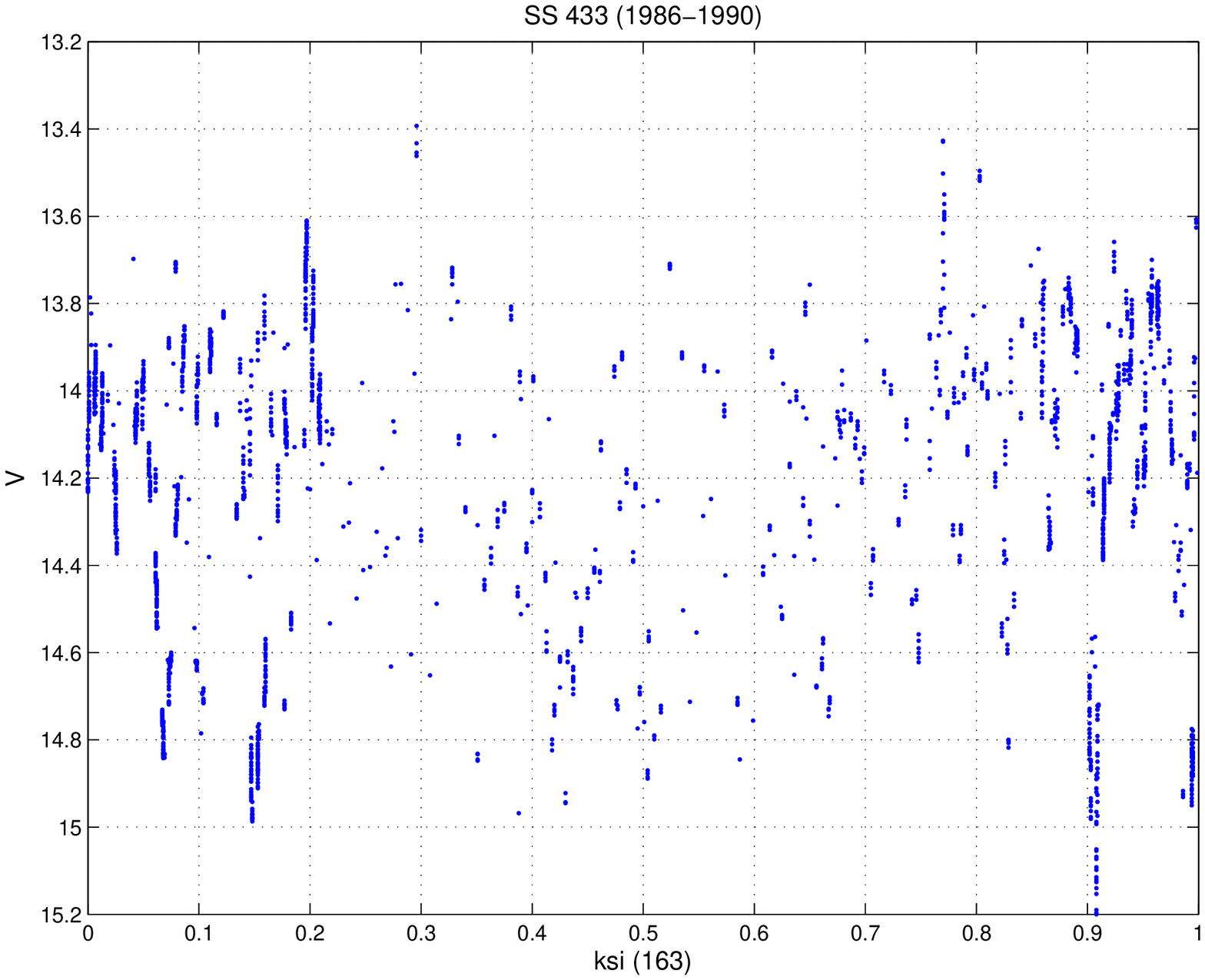,width=150mm}}
\caption{The 1986-1990 V-band light curve folded with the
precession phase}
\end{figure}

\newpage
\begin{figure}[b]
\centerline{\epsfig{file=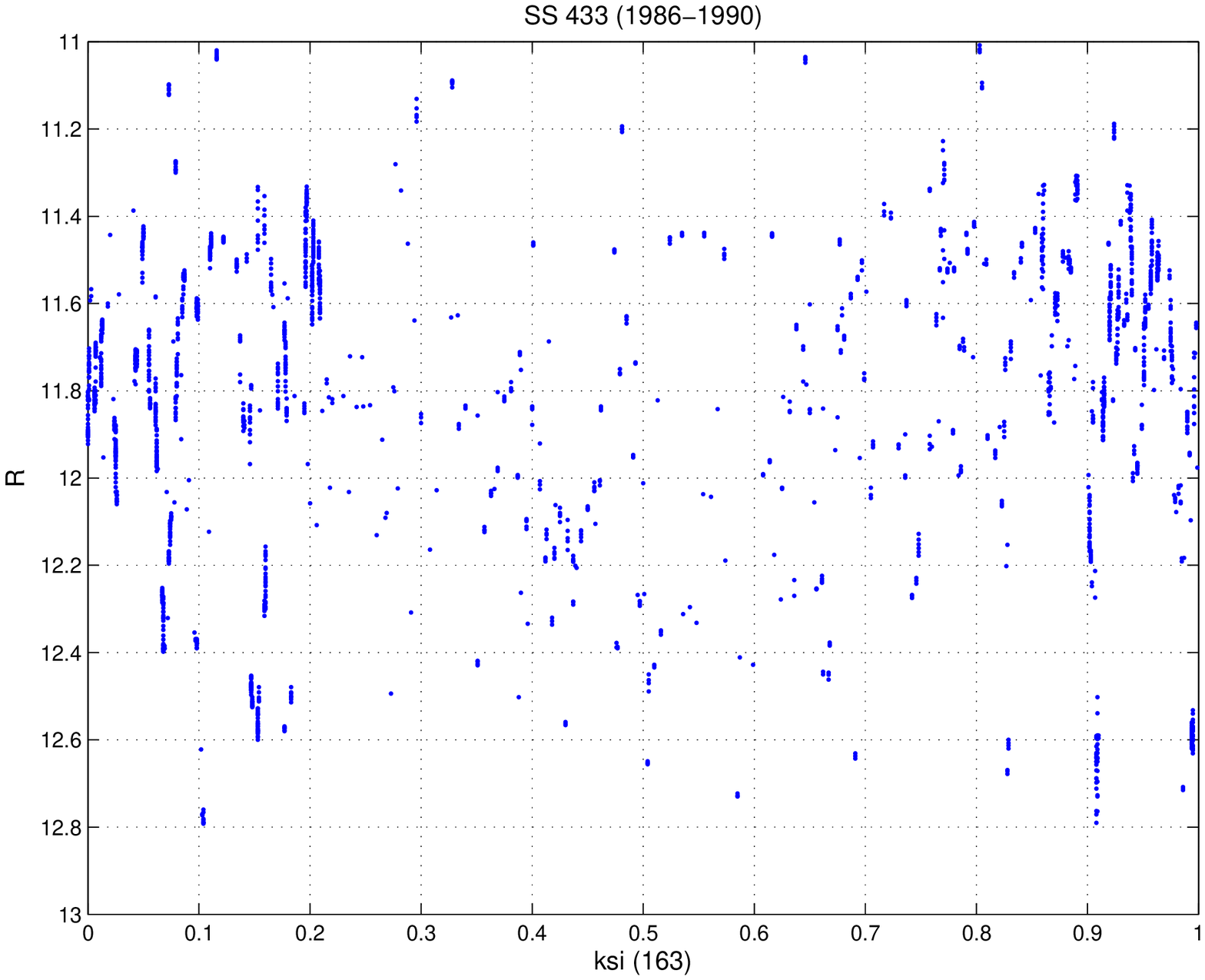,width=150mm}}
\caption{The 1986-1990 R-band light curve folded with the
precession phase}
\end{figure}

\newpage
\begin{figure}[b]
\centerline{\epsfig{file=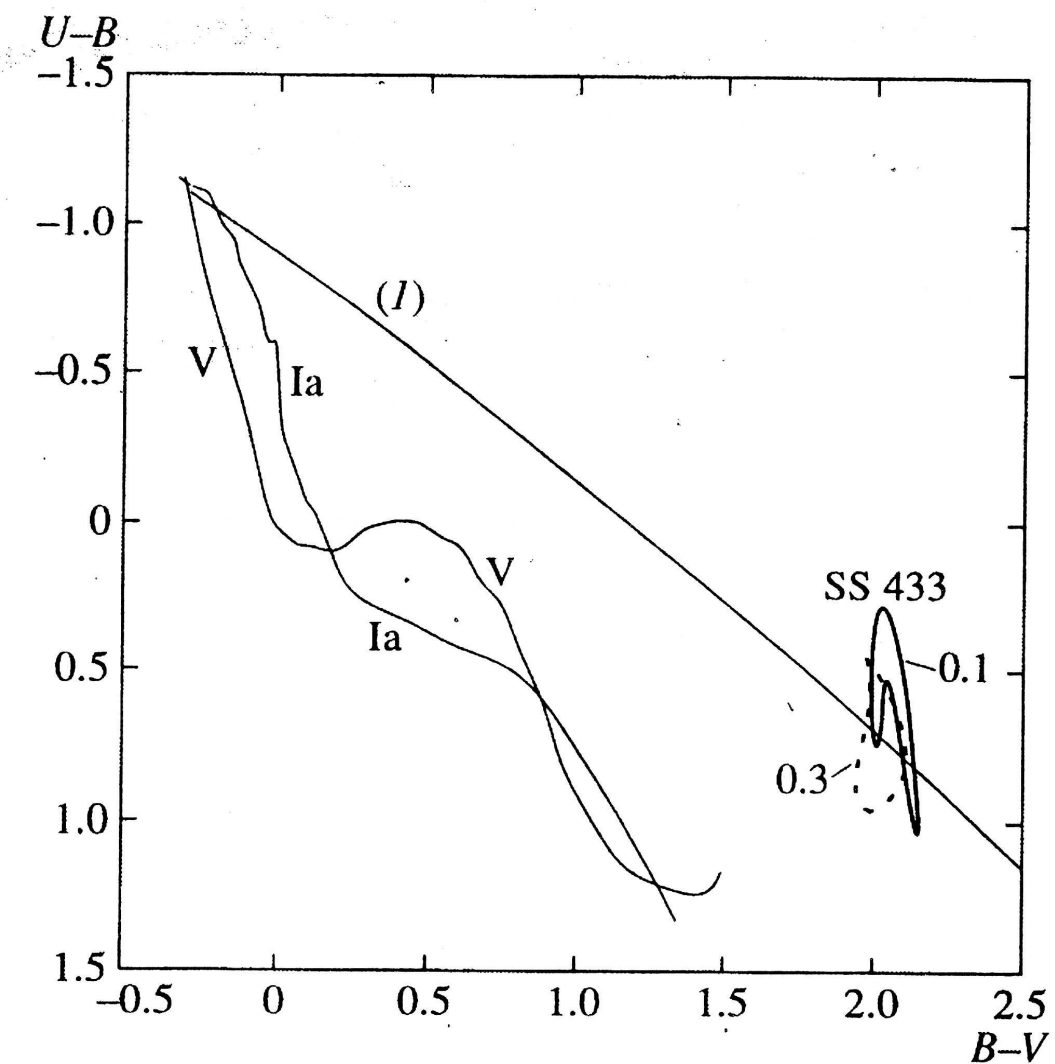,width=150mm}}
\caption{Smoothed orbital tracks of SS 433 on two-color diagram
U-B, B-V. Mean precession phases are indicated by the numbers.
Normal color curves for dwarfs and supergiants and reddening line
for not stars are shown by the solid lines.}
\end{figure}

\end{document}